\documentstyle[twocolumn,floats,pre,aps]{revtex}

\renewcommand{\usepackage}{\RequirePackage}
\usepackage{times}
\usepackage{graphicx}
\usepackage{psfrag}

\tighten
\DeclareMathAlphabet{\mathrmb}{OT1}{cmr}{b}{n}
\DeclareMathAlphabet{\mathsfb}{OT1}{cmss}{bx}{n}

\renewcommand\ensuremath\relax
\newcommand\eqref[1]{(\ref{#1})}

\newcommand{\xlabel}[1]{\psfrag{xlabel}[t][]{\small #1}}
\newcommand{\ylabel}[1]{\psfrag{ylabel}[B][]{\small #1}}

\newcommand{\gpfig}[1]{%
  \resizebox{8.6cm}{!}{\includegraphics{#1}}\\[\baselineskip]
  }

\newcommand{\pder}[2]{\ensuremath{\frac{\partial#1}{\partial#2}}}
\newcommand{\Int}[3]{\int\limits_{#1}^{#2}\!\!\!{d}#3\,}

\newcommand{\braket}[1]{\left\langle#1\right\rangle}

\newcommand{\Fig}[1]{Fig.~\ref{#1}}
\newcommand{\Figs}[1]{Figs.~\ref{#1}}
\newcommand{\Eq}[1]{Eq.~\eqref{#1}}

\newcommand{\ndp}{N \sigma}
\newcommand{\bu}{{\bf u}}

\newcommand{\mt}{}
\newcommand{\pmt}{}
\newcommand{\pers}{}
\newcommand{\mps}{}
\newcommand{\mpss}{}

\title{Temperature scaling in a dense vibro-fluidised granular material}
\author{P Sunthar and V Kumaran}
\address{Department of Chemical Engineering, \\ 
  Indian Institute of Science, Bangalore, 560 012, India.}

\begin{document}

\maketitle

\begin{abstract}
  The leading order ``temperature'' of a dense two dimensional granular
  material fluidised by external vibrations is determined.  The grain
  interactions are characterised by inelastic collisions, but the
  coefficient of restitution is considered to be close to $1$, so that
  the dissipation of energy during a collision is small compared to
  the average energy of a particle. An asymptotic solution is obtained
  where the particles are considered to be elastic in the leading
  approximation. The velocity distribution is a Maxwell-Boltzmann
  distribution in the leading approximation. The density profile is
  determined by solving the momentum balance equation in the vertical
  direction, where the relation between the pressure and density is
  provided by the virial equation of state. The temperature is
  determined by relating the source of energy due to the vibrating
  surface and the energy dissipation due to inelastic collisions. The
  predictions of the present analysis show good agreement with
  simulation results at higher densities where theories for a dilute
  vibrated granular material, with the pressure-density relation
  provided by the ideal gas law, are in error.
\end{abstract}

\section{Introduction}
Recent developments in the physics of granular matter \cite{jae-nag96}
have illustrated that the dissipative nature of the interactions
between grains can result in a variety of different phenomena. Of
particular interest in recent years has been the dynamics of vibrated
granular materials \cite{warretal95,meloetal95}, which exhibit
stationary states as well as waves and complex patterns. In order to
describe these diverse states of the material, it is necessary to
derive macroscopic descriptions by averaging over the microscopic
details of the motion and interactions between individual grains.
This goal has proved elusive, however, because a vibrated granular
material is a driven dissipative system, and the interactions between
the particles are characterised by a loss of energy due to inelastic
collisions. The statistical mechanics framework developed for
equilibrium or near equilibrium systems cannot be used in this case.
Consequently, phenomenological models
\cite{shrin97,tsim-aran97,venkat-ott98} have been used to describe the
dynamics of granular materials. The kinetic theories developed for
granular flows \cite{jen-sav83,kum98:vib} usually assume that the
system is close to ``equilibrium'' and the velocity distribution
function is close to the Maxwell-Boltzmann distribution.

Experimental studies and computer simulations have reported the
presence of a uniformly fluidised state in a vibrated bed of granular
material.  Luding, Herrmann and Blumen \cite{ludetal94} carried out
`Event Driven' (ED) simulations of a two dimensional system of
inelastic disks in a gravitational field vibrated from below, and
obtained scaling laws for the density variations in the bed. An
experimental study of a vibrated fluidised bed was carried out by
Warr, Huntley and Jacques \cite{warretal95}.  Their experimental set
up consisted of steel spheres confined between two glass plates that
are separated by a distance slightly larger than the diameter of the
spheres. The particles were fluidised by a vibrating surface at the
bottom of the bed, and the statistics of the velocity distribution of
the particles were obtained using visualisation techniques. Profiles
for the density and the mean square velocity were obtained, and the
particle velocity distributions were also determined at certain
positions in the bed.  Both of these studies reported that there is an
exponential dependence of the density on the height near the top of
the bed, similar to the Boltzmann distribution for the density of a
gas in a gravitational field. However, the dependence of the density
deviates from the exponential behaviour near the bottom. The
dependence of the mean square velocity on the vibration frequency
and amplitude were found to be different in the two studies.

A theoretical calculation of the distribution function in a
vibro-fluidised bed was carried out by Kumaran
\cite{kum98:vib,kum98:vibscal}. The limit of low dissipation, where
the coefficient of restitution $e$ is close to $1$ was considered. In
this limit, the mean square velocity of the particles is large
compared to the mean square of the velocity of the vibrating surface,
and the dissipation of energy during a binary collision is small
compared to the energy of a particle. A perturbation approximation is
used, where the energy dissipation is neglected in the leading order
approximation, and the system resembles a gas at equilibrium in a
gravitational field. The velocity distribution function is a
Maxwell-Boltzmann distribution, and the density decreases
exponentially from the vibrating surface.  The first order correction
to the distribution due to dissipative effects was calculated using
the moment expansion method, and the results were found to be in
qualitative agreement with the experiments of Warr et. al.
\cite{warretal95}.

The theoretical predictions \cite{kum98:vib,kum98:vibscal} were
compared with previous experimental and simulation studies by McNamara
and Luding \cite{mclud98}. They found that the theory was in good
agreement with experiments for dilute beds, where the area fraction of
the particles is low, but there were systematic deviations from the
theoretical predictions as the area fraction increases. This is to be
expected, since the analysis assumed that the density is small and the
pair distribution function was set equal to $1$ and therefore the
pressure is related to the density by the ideal gas law. These
assumptions become inaccurate as the area fraction of the bed
increases. An approximate method for including the correction to the
pair distribution function was suggested by Huntley \cite{hunt98}.

In the present analysis, the correction to the low density theory of
Kumaran \cite{kum98:vib,kum98:vibscal} is determined for a
vibro-fluidised bed where the coefficient of restitution is close to
$1$. An asymptotic analysis is used, where the dissipation is
neglected in the leading approximation. The leading order density and
velocity profiles are determined using the momentum balance equation
in the vertical direction. In contrast to the earlier theory
\cite{kum98:vib,kum98:vibscal}, the virial equation of state for a
non-ideal two dimensional gas is used to determine the leading order
density profile. The density profile differs from the Boltzmann
distribution, but the velocity distribution function is still a
Maxwell-Boltzmann distribution. The leading order temperature is
determined by a balance between the source and dissipation of energy
as before. The complete equilibrium pair distribution function is used
to determine the rate of dissipation of energy due to inelastic
collisions. The results are compared with hard sphere MD
simulations, and also with earlier theoretical and simulation studies.

\section{Analysis}
The system consists of a bed of circular disks (of diameter $\sigma$) in
a gravitational field driven by a vibrating surface. The vibrating
surface has a periodic amplitude function but no assumption is made
regarding the form of the function. There is a source of
energy at the vibrating surface due to particle collisions with the
surface, and the dissipation is due to inelastic collisions. A balance
between the two determines the ``temperature'', which is the mean
square velocity of the particles.

The limit of low dissipation, where the coefficient of restitution $e$
is close to $1$, is considered. In this limit, it can be shown that
the mean square velocity of the particles is large compared to the
mean square velocity of the vibrating surface.  An asymptotic
expansion in the parameter $\epsilon \equiv U_0^2/T_0$ is used
\cite{kum98:vib}. If the source and dissipation of energy are
neglected in the leading approximation, the system resembles a gas of
hard disks at equilibrium in a gravitational field.  The velocity
distribution function is a Maxwell-Boltzmann distribution at
equilibrium
\begin{equation}
  F(\bu) = \frac{1}{2 \pi T_{0}} \exp{ \left( - \frac{u^{2}}{2 
        T_{0}} \right)},
\end{equation}
where $T_{0}$ is the leading order temperature. The density profile is
determined by solving the momentum balance equation in the vertical
direction
\begin{equation}
  \label{eq:mombal}
  \pder{p}{z} - \rho g = 0,
\end{equation}
where $p$ is the pressure, $\rho$ is the density (number of particles
per area) and $g$ is the acceleration due to gravity. For a gas at
equilibrium, the pressure is related to the density by the virial
equation of state, which in the case of inelastic circular disks is
\begin{equation}
  p = \rho T_{0} \left[\frac{1 + e}{2} + (1 + e) g_0(\nu) \, \nu \right],
\end{equation}
where $g_{0}(\nu)$ is the pair distribution function at contact,
which for circular disks is given by \cite{verlet82}
\begin{equation}
  g_{0}(\nu) = \frac{1}{16 (1 - \nu)^{2}} \left[ 16 - 7 \nu - 
    \frac{\nu^{3}}{4 (1 - \nu)^{2}} \right],
\end{equation} 
and $\nu$ is the area fraction corresponding to $\rho$.  If the
coefficient of restitution is set equal to $1$ in the leading
approximation, the equation for the pressure reduces to the standard
virial equation of state
\begin{equation}
  p = \rho T_{0} \left[ 1 + 2 g_{0}(\nu) \, \nu\right].
\end{equation}
The resulting equation from \Eq{eq:mombal} for the density profile is
a first order ordinary differential equation, which can be solved
using the mass conservation condition
\begin{equation}
  \label{eq:masscons}
  \Int{0}{\infty}{z} \rho = N,
\end{equation}
where $N$ is the number of particles per unit width of the bed.  Note
that the leading order temperature $T_{0}$ is still unknown at this
stage. This is determined using a balance between the source and
dissipation of energy.  The source of energy due to particle
collisions with the vibrating surface is determined using an
equilibrium average over the increase in energy due to particle
collisions with the vibrating surface \cite{kum98:vib,kum98:vibscal}
\begin{equation}
  \label{eq:s0}
  S_{0} = 2 \sqrt{\frac{2}{\pi}} \, T_{0}^{1/2}
  \braket{U^2} \, g_0(\nu) \, \rho \, \Big|_{z=0}.
\end{equation}
Here $\braket{U^2}$ represents the mean square velocity of the
vibrating surface.  The rate of dissipation of energy per unit width
is calculated by averaging over the energy loss over all the
collisions between particles and integrating over the height of the
bed \cite{kum98:vib}
\begin{equation}
  \label{eq:d0e}
  D_0 =  \sqrt{\pi} \, \sigma (1-e^2)\, T_0^{3/2}  
  \Int{0}{\infty}{z} g_0(\nu) \, \rho^2.
\end{equation}
Note that the $g_0$ appearing in $S_0$ and $D_0$ is the Enskog factor
which accounts for the increase in the frequency of collision for hard
disks at high densities.  The temperature $T_{0}$ can now be
determined from the relation
\begin{equation}
  \label{eq:s0d0}
  S_0 = D_0
\end{equation}
An analytical solution to the density variation \Eq{eq:mombal} can be
determined in the low density limit using the equation of state for an
ideal gas for the pressure \cite{kum98:vib}.
\begin{equation}
  \label{eq:nulow}
  \rho = \frac{N g}{T_{0}} \exp{ \left( - \frac{g z}{T_{0}} 
\right)}
\end{equation}
where the leading order temperature is given by,
\begin{equation}
  T_{0} = \frac{4 \sqrt{2}}{\pi} \frac{\braket{U^2}}{N \sigma (1 -
  e^{2})}.
\end{equation}
In the low-density limit the density decays exponentially from the
bottom of the bed. At higher densities the solution to the density
variation is no longer exponential throughout, and has to be obtained
numerically by an iterative scheme.  However, at large distances from
the bottom, the bed is dilute and the ideal gas law holds good, hence
the decay is exponential, even though near the bottom it is not. This
gives a convenient starting point for the numerical integration from a
\emph{finite} height, above which we assume the asymptotic solution
($z \rightarrow \infty$) to be given by an exponential decay known to
within two undetermined constants.  A value for the density and the
temperature is assumed at this height and the integration is carried
out up to the vibrating plate ($z=0$).  The complete density profile
is obtained by combining the numerical and the asymptotic solutions.
If the conditions \Eq{eq:masscons} and \Eq{eq:s0d0} are not satisfied
after one such integration, a new value is determined for the density
and temperature using the Newton-Raphson method, and the iteration is
repeated till convergence. In cases where the convergence is poor, the
solution is obtained by \emph{continuing} a low density solution in a
parameter such as $N \sigma$ or $U_0$.

\textbf{Viscous dissipation:} The above analysis can be easily
extended to the case of dissipation purely due to viscous drag. The
expression for the source of energy remains the same as given by
\Eq{eq:s0}. A drag law given by $a_i = -\mu u_i$ is assumed. The total
leading order rate of dissipation per unit width will then be 
\begin{eqnarray}
  \label{eq:d0v}
  D_{D0} & = & \mu \Int{0}{\infty}{z} \, \rho \Int{}{}{\bu} \, F(\bu)
             \,\, \bu\cdot\bu \nonumber\\
      & = & 2 \mu N T_0
\end{eqnarray}
Unlike \Eq{eq:d0e}, the leading order dissipation is the same for the
low density and the high density cases. Nevertheless, the density profile
has to be obtained numerically in the manner outlined above, with
\Eq{eq:d0v} substituted for \Eq{eq:d0e} in \Eq{eq:s0d0}.

\section{Simulation and Results}
The hard sphere molecular dynamics (MD), also known as event driven
(ED) method \cite{ludetal94} is used for the simulations of
the vibro-fluidised bed. Periodic boundary conditions are used in the
horizontal direction and the vibrating surface at the bottom has a
sawtooth form for the amplitude function. The simulations are carried
out only for the case of inelastic collisions, since the viscous drag
requires a different treatment than the ED method. 

\begin{figure}[tb]
  \begin{center}
    \xlabel{$z/\sigma$}
    \ylabel{$\nu$}
    \gpfig{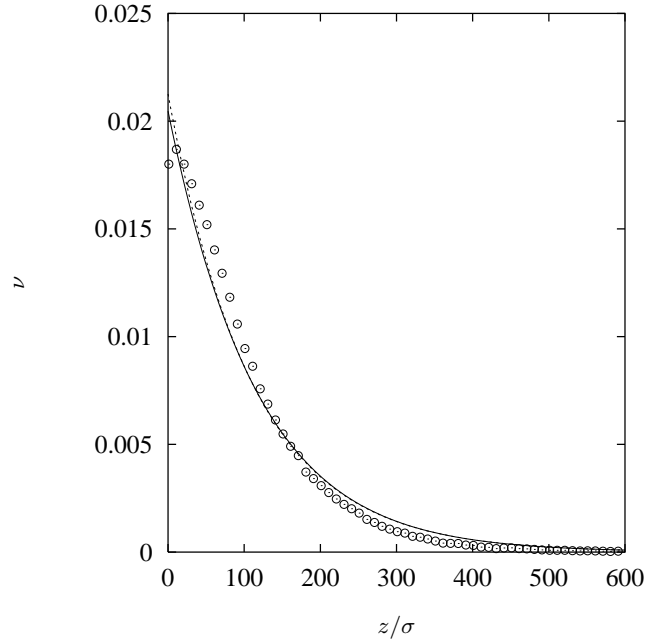}
    \caption{Exponential decay of packing fraction ($\nu$) with
      a normalised height ($z/\sigma$) at low densities. The
      predictions of the present analysis (solid line) and the low
      density theory (dotted line) of \protect\cite{kum98:vib} is
      compared with simulation (points).  Both the predictions are
      nearly identical. Here, $\epsilon=0.3$, $\ndp=3$, $g=1\mpss$, and
      $U_0=6\mps$.}
    \label{fig:ldlenu}
  \end{center}
\end{figure}

\begin{figure}[tb]
  \begin{center}
    \xlabel{$z/\sigma$}
    \ylabel{$\nu$}
    \gpfig{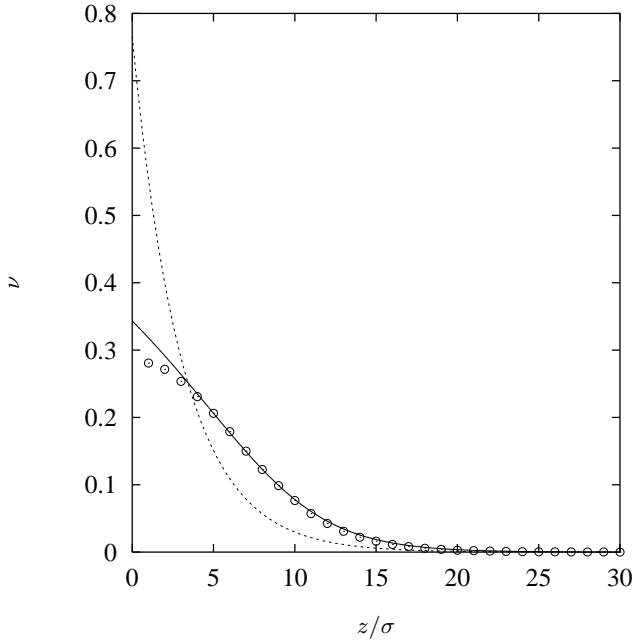}
    \caption{Deviation of the density profile from the exponential decay
      at high densities in the case of dissipation due to inelastic
      collisions. The simulation result (points) is captured by the
      present analysis (solid line) which is lower than the
      exponential decay (dotted line) of the low density theory of
      \protect\cite{kum98:vib} near the bottom of the bed. Here
      $\epsilon=0.3$, $\ndp=3$, $g=1\mpss$, and $U_0=1\mps$.}
    \label{fig:hdlenu}
  \end{center}
\end{figure}

\begin{figure}[tb]
  \begin{center}
    \xlabel{$z/\sigma$}
    \ylabel{$\nu$}
    \gpfig{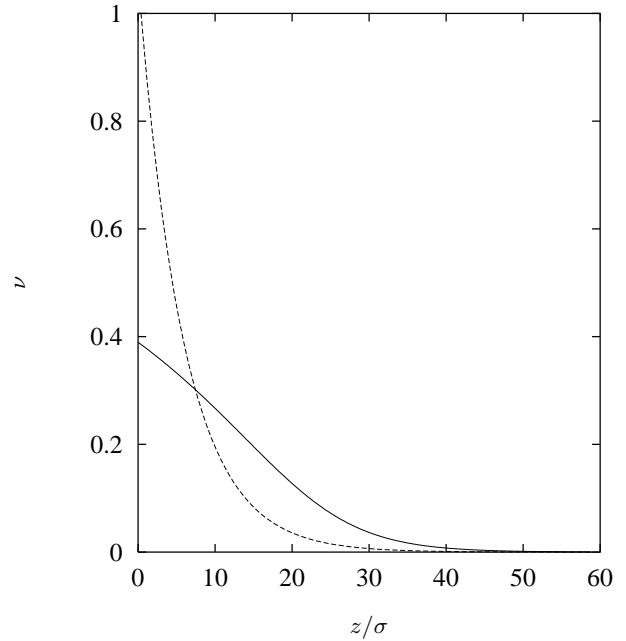}
    \caption{Deviation of the density profile from the exponential decay
      at high densities in the case of dissipation due to viscous
      drag. The present analysis (solid line) gives physically
      plausible values for the packing fraction near the bottom, while
      the low density theory (dotted line) of \protect\cite{kum98:vib}
      predicts values higher than the maximum closed packing. Here
      $\epsilon=0.2$, $\ndp=20$, $g=20\mpss$, $\mu=0.1\pers$, and
      $U_0=5\mps$.}
    \label{fig:hvis}
  \end{center}
\end{figure}

The density profiles obtained using the present analysis, as well as
the earlier low density approximations of Kumaran \cite{kum98:vib},
are compared with the simulation results in \Figs{fig:ldlenu}
and~\ref{fig:hdlenu}.  It is seen that the density profiles of the
present analysis are in good agreement with the simulation results
even when the density near the bottom of the bed becomes large, while
the profiles from the low density approximation have significant
errors. \Fig{fig:hvis} shows the nature of the density profile in the
high density limit in the case of dissipation due to viscous drag.
Here too the present analysis gives reasonable values for packing
fraction near the bottom, while the low density theory predicts
physically incorrect values. 

In a recent work, McNamara and Luding \cite{mclud98} reported the
scaling of dissipation with the center of mass obtained from
simulations. The results agreed with the low density theory of
\cite{kum98:vibscal} but a systematic deviation was observed at
high densities in all the cases. This deviation is captured in the
present analysis.  The leading order dissipation at low densities in
the bed is given by \cite{kum98:vib}
\begin{equation}
  \label{eq:d0}
  D_0 = \frac{\sqrt{\pi}}{2}\, (1-e^2) \, N^2 \sigma g \sqrt{T_0}.
\end{equation}
In \cite{mclud98} the total dissipation obtained from the simulation
was normalised by a factor taken out from this leading order
dissipation and a non dimensional number was defined as
\begin{equation}
  \label{e:cpp}
  C_{pp} \equiv  \frac{D_0}{(1-e) N^2 \sigma g \sqrt{T_0/2}}.
\end{equation}
The scaling of this factor with the height of the center of mass ($h$)
above the position at rest ($h_0$) was studied.  This factor was found
out for different parameter sets by varying the bottom wall velocity
$U_0$ over several decades such that the bed is taken from a densely
packed regime to a very low density regime. They chose a central
data set and varied the parameters one at a time.  It was found that
in all the cases considered, the scaling relation collapsed to a
single curve. The central parameter set has the following values $N =
3.2$, $\sigma=1$, $g=1$, $e=0.95$.  

The present analysis is valid when $\epsilon \equiv U_0^2/T_0 \ll 1$
and when the frequency of particle-particle collision is much greater
than the frequency of particle-wall collisions. It can be shown that
in the leading order the ratio of the frequency of particle-particle
collisions to the frequency particle-wall collisions is $ \sqrt{2}\pi
\,\ndp$.  Hence the present analysis will hold good when $\ndp \gg
1/\sqrt{2}\pi$.  The central set corresponds to $\epsilon = 0.35$,
$\ndp = 3.2$ and therefore we expect the present analysis to hold good
for this case.  Most of the parameter sets used in \cite{mclud98} also
fall within the limits of the theory derived here.

\begin{figure}[tb]
  \begin{center}
    \xlabel{$2(h-h_0)/\sigma$}
    \ylabel{$C_{pp}$}
    \gpfig{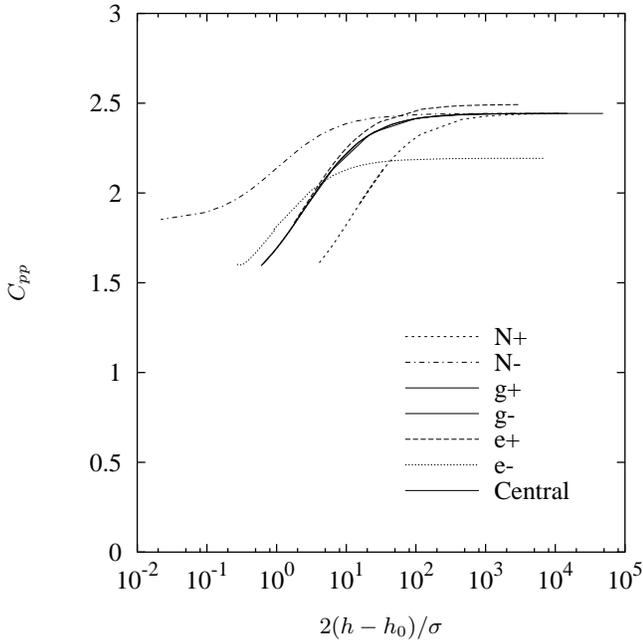}   
    \caption{Theoretical scaling of the normalised dissipation
      ($C_{pp}$) against the center of mass ($h$) above the position
      at rest ($h_0$) for the different cases reported in
      \protect\cite{mclud98}. All except two---(N+) with $\epsilon =
      1.73$ and (N--) with $\ndp = 0.65$ collapse on to a single curve
      in the linear region. The parameters indicated correspond to
      $N=16 \pmt$ (N+), $N=0.65\pmt$ (N--), $g=25\mpss$ (g+),
      $g=0.04\mpss$ (g--), $e=0.99$ (e+), $e=0.75$ (e--), rest of the
      parameters being same as the one in the central set, which has
      the following values $N = 3.2\pmt$, $\sigma=1\mt$, $g=1\mpss$,
      $e=0.95$.}
    \label{fig:theoscal}
  \end{center}
\end{figure}

\begin{figure}[tb]
  \begin{center}
    \xlabel{$2(h-h_0)/\sigma$}
    \ylabel{$C_{pp}$}
    \gpfig{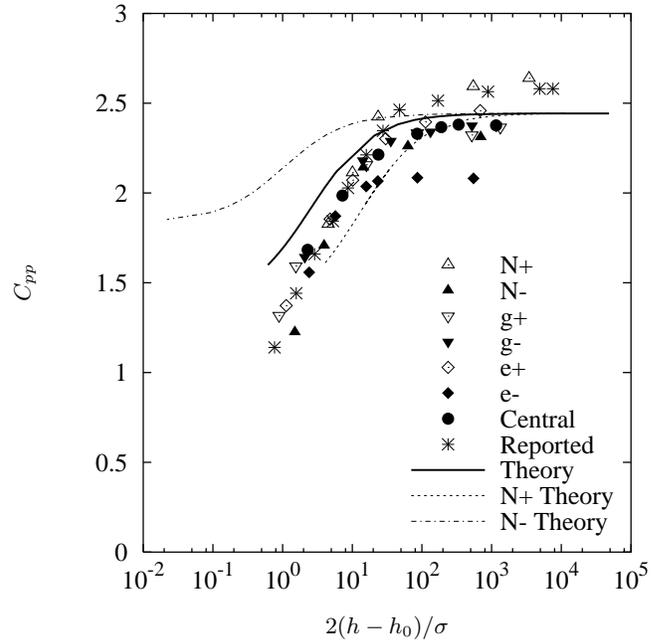}
    \caption{Scaling of the normalised dissipation with the center 
      of mass: Predictions from the present analysis is compared with
      the results from our simulations and the reported results in
      \protect\cite{mclud98}.  The linear portion of all the curves
      from theory, except two, fall on the solid line denoted as
      `Theory'. The two exceptions are also shown.  A set of points
      correspond to the simulation data with parameter values $N =
      16\pmt$ (N+), $N=0.65\pmt$ (N--), $g=25\mpss$ (g+),
      $g=0.04\mpss$ (g--), $e=0.99$ (e+), $e=0.75$ (e--); rest of the
      parameters in a set being the same as the one in the central
      set, which has the following values $N = 3.2\pmt$, $\sigma=1\mt$,
      $g=1\mpss$, $e=0.95$.}
    \label{fig:simtheo}
  \end{center}
\end{figure}

\Fig{fig:theoscal} shows the theoretical predictions of the total
dissipation for the different cases reported in Fig.~2 in
\cite{mclud98}.  It is compared with the results of two simulations in
\Fig{fig:simtheo}. It is seen that the present analysis correctly
predicts the lowering of the coefficient $C_{pp}$ at high densities.
This reduction in the dissipation from the constant value at low
densities is the net result of two opposing factors: (i) decrease in
the density from the exponential behaviour near the vibrating bottom
(see \Fig{fig:hdlenu}), hence reducing the total value of the
dissipation, and (ii) increase in frequency of collisions at high
densities, increasing the dissipation.  

It is also seen that not all the theoretical predictions collapse on
to a curve as is the case with the data from the simulation. In two of
the cases the theory does not agree with the simulations because (i)
in one the value of the perturbation parameter is high ($\epsilon =
1.73$) and the leading order theory is valid only for low $\epsilon$,
and (ii) in the other case the value of $\ndp = 0.65 $ is low. 

In \Fig{fig:theoscal}, the apparent mismatch with `e-' is not a
discrepancy with the model, but has got to do with the formula chosen
used in \cite{mclud98} for the normalisation of the dissipation factor
$C_{pp}$. They had chosen to normalise the dissipation by a factor
$(1-e)$. While this might have given a better fit for high densities
(low center of mass), the correct factor for very low densities is
$(1-e^2)$ as given by \Eq{eq:d0}.  The difference is more pronounced in
the case of $e\ll 1$, which, here, has a value $e=0.75$. A close
inspection of the curves `e-' in \Fig{fig:theoscal} and
\Fig{fig:simtheo} show that the theory and simulation do indeed agree
with each other.

We also note here that the data taken from the reported simulation
\cite{mclud98} is for asymmetric sawtooth vibration, whereas our
simulation is for the symmetric sawtooth. Both these give similar
results for the scaling of $C_{pp}$. Also the theoretical predictions
for the symmetric and the asymmetric sawtooth are identical,
indicating that the form of the bottom wall vibration does not affect
the scaling of the dissipation with the center of mass.

\section{Conclusion}
In summary, a theory to describe the state of a vibro-fluidised bed in
the dense limit was derived. This is different from the earlier theory
of Kumaran \cite{kum98:vib,kum98:vibscal}, which is valid for low
densities where the ideal gas equation was used and the pair
distribution function was set equal to $1$. We have made use of the
virial equation of state to obtain a correction to the exponential
density profile obtained in low densities and the pair distribution
function is used to calculate the increased frequency of collisions in
the source and the dissipation of energy.  The theoretical predictions
of density and temperature were compared with the results obtained
from MD simulation of two dimensional disks.  The theory correctly
predicts the lowering of the density from the exponential value at
high densities near the bottom.  The theory also predicts the scaling
relations of the total dissipation in the bed reported in
\cite{mclud98}.


\begin{thebibliography}{10}

\bibitem{jae-nag96}
H.~M. Jaeger and S.~R. Nagel, Rev. Mod. Phys. {\bf 68},  1259  (1996).

\bibitem{warretal95}
S. Warr, J.~M. Huntley, and G.~T.~H. Jackques, Phys. Rev. E {\bf 52},  5583
  (1995).

\bibitem{meloetal95}
F. Melo, P.~B. Umbanhowar, and H.~L. Swinney, Phys. Rev. E {\bf 75},  3838
  (1995).

\bibitem{shrin97}
T. Shrinbot, Nature {\bf 383},  574  (1997).

\bibitem{tsim-aran97}
S. Tsimring and I.~S. Aranson, Phys. Rev. Lett. {\bf 79},  213  (1997).

\bibitem{venkat-ott98}
S.~C. Venkataramani and E. Ott, Phys. Rev. Lett. {\bf 80},  3495  (1998).

\bibitem{jen-sav83}
J.~T. Jenkins and S.~B. Savage, J. Fluid Mech. {\bf 130},  187  (1983).

\bibitem{kum98:vib}
V. Kumaran, J. Fluid Mech. {\bf 364},  163  (1998).

\bibitem{ludetal94}
S. Luding, H.~J. Herrmann, and A. Blumen, Phys. Rev. E {\bf 50},  3100  (1994).

\bibitem{kum98:vibscal}
V. Kumaran, Phys. Rev. E {\bf 57},  5660  (1998).

\bibitem{mclud98}
S. Mc{N}amara and S. Luding, Phys. Rev. E {\bf 58},  813  (1998).

\bibitem{hunt98}
J.~M. Huntley, Phys. Rev. E {\bf 58},  5168  (1998).

\bibitem{verlet82}
L. Verlet and D. Levesque, Mol. Phys. {\bf 46},  969  (1982).

\end{thebibliography}

\end{document}